\documentclass{calor2002}

\begin{document}

\title{Calorimeters in Astro and Particle physics}

\author{Klaus Pretzl}

\address{Laboratory for High Energy Physics, University of Bern, \\ 
3012 Bern, Switzerland\\ 
E-mail: pretzl@lhep.unibe.ch}


\maketitle

\abstracts{
In this article an attempt is made to review some of the original works leading to new developments of calorimeters which are so widely and successfully used in astro and particle physics experiments. This report is far from being complete and the author apologizes for  omissions and misquotations.}

\section{Introduction}

Calorimeters belong to the most important instruments to measure the energy of neutral and charged particles produced with cosmic rays or with particle accelerators. They provide the means to explore new physics in an energy range from several eV to more than $10^{20}$ eV. Their development was very much driven by the quest for new frontiers in astro- and particle physics.

Several types of calorimeters have been developed. There are the so-called true calorimeters, which operate at very low temperatures and are used as thermal sensors. Cryogenic calorimeters are the most sensitive devices in the energy region eV to keV. They are frequently used in direct dark matter detection and double beta decay experiments. Then there are the sampling calorimeters, which are based on measuring the energy loss of secondary particles in an active material, which is inserted between the absorbing material of the calorimeter (see for example Fig.1). Ionization or scintillation detectors are commonly used as active layers. Sampling calorimeters are among the most frequently employed calorimeters in high energy physics experiments because they are relatively cheap and they can cover a large energy range, typically from GeV to TeV. Homogeneous total absorption calorimeters, like crystals or liquids (Ar, Kr, Xe) have mostly been developed for electromagnetic shower detection. They yield outstanding energy resolutions but they are rather expensive. Cerenkov calorimeters are based on the detection of Cerenkov radiation of relativistic particles in transparent materials. An early example is the lead glass calorimeter for electromagnetic shower detection.

Cerenkov calorimeters with enormous dimensions using large underground water tanks, sea water, polar ice and the atmosphere of the earth have recently been developed to measure solar, atmospheric and cosmic neutrinos as well as ultra high energy particles (UHEP) in cosmic rays. Large atmospheric calorimeters also make use of the fluorescence scintillation light emitted when particles pass 
through the atmosphere. The observation of the fluorescence light of air showers in the atmosphere from a space station is planned for the near future. Properly instrumented, the atmosphere would be the largest calorimeter ever put in operation and would enable us to study UHEP with energies in excess of $10^{20}$ eV.

There are excellent review articles\cite{bib:gian} as well as a monograph\cite{bib:wigm}  which describe in all details the state of the art in calorimetry. It is the intention of this article to highlight some of the original works in the development of calorimeters and their impact on experiments in astro and particle physics.

\section{Early Developments}

One of the early pioneers in developing the art of calorimetry was W. Orthmann, a close collaborator of W. Nernst (Nobel prize winner for chemistry in 1920). Orthmann\cite{bib:ort} developed a differential calorimeter with which he could measure heat transfers of the order of $\mu\mathrm{Watt}$. Using this true calorimetric technique, he and L. Meitner\cite{bib:mei} were able to determine the mean energy of the continuous $\beta\mathrm{-spectrum}$ in $^{210}\mathrm{Bi}$ to be E = 0.337 MeV $\pm 6\%$. This value agrees very well with the mean kinetic energy of E = 0.33 MeV of the emitted $\beta\mathrm{-particles}$. Their measurements contributed to the notion of a continuous $\beta\mathrm{-spectrum}$ leading to Pauli's neutrino hypothesis in 1930.

Applying this technique to high energy particles would fail, since the temperature increase $\Delta\mathrm{T}=\frac{\Delta\mathrm{E}}{C}$ caused by the energy loss $\Delta\mathrm{E}$ of a high energy particle is unmeasurably small ($\Delta\mathrm{T}\sim10^{-7}\mathrm{K}$ for $\Delta\mathrm{E}\sim10^{20}$ eV) due to the heat capacity of the large absorber mass ($\sim10^{8}$ g Fe) necessary to contain the shower. Nevertheless this technique was revitalized in the 1980s by the development of calorimeters for dark matter detection (see chapter 9).

\begin{figure}[th]
\centerline{\epsfxsize=3.2in\epsfbox{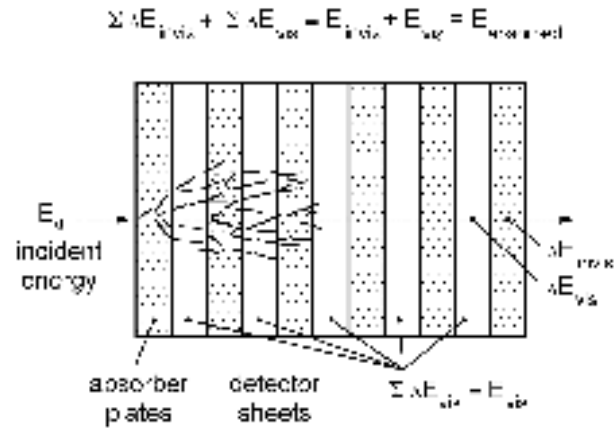}}   
\caption{The principle of a sampling calorimeter is shown. \label{fig:inter}}
\end{figure}

\begin{figure}[th]
\centerline{\epsfxsize=4in\epsfbox{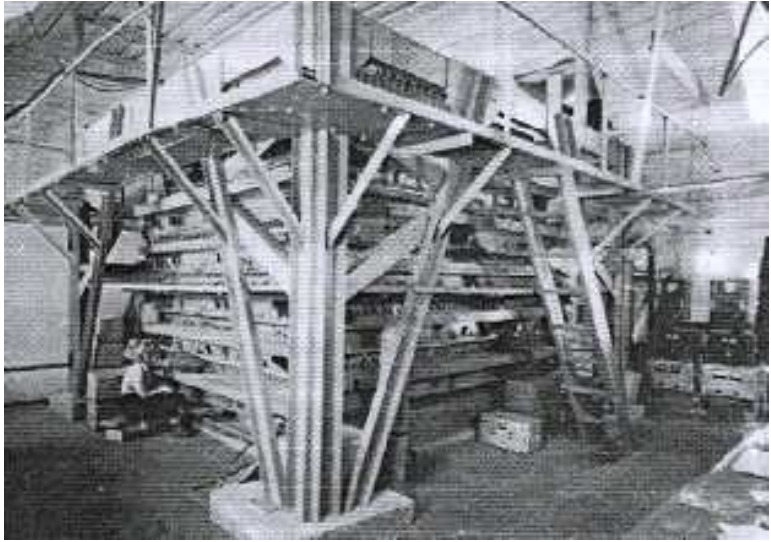}}   
\caption{The ionization sampling  calorimeter of Babayan, Grigorov, Shestoperov and Sobikyakov used in cojunction with nuclear emulsions. \label{fig:inter2}}
\end{figure}

In 1954 N.L. Grigorov\cite{bib:gri} put forward the idea of sampling calorimeters using ionisation chambers (arrays of proportional counters) interspersed between thick iron absorber sheets to measure cosmic ray particles with energies $E>10^{14}$ eV. The visible energy of the particle  $E_{visible}=\frac{\mathrm{dE}}{\mathrm{dx}}\int\mathrm{n(x)dx}$ is then determined from the number of secondary particles in the shower n(x) and their energy loss $\frac{\mathrm{dE}}{\mathrm{dx}}$ in the ionisation detectors.
 Because of the invisible energy loss in the absorber plates and in the detector sheets, sampling calorimeters need to be calibrated in particle beams with known energies in order to obtain absolute energy measurements. The principle of a sampling calorimeter is shown in Fig.\ref{fig:inter}.
In 1957 Grigorov and his collaborators\cite{bib:mur} constructed a sampling calorimeter in the Pamir mountains, at an altitude of 3860 m above sea level (Fig.\ref{fig:inter2}). Their calorimeter also employed $10 m^{2}$ of emulsion sheets, which they placed between 3 lead sheets to study details of the primary interaction. In order to identify corresponding events in the emulsions and the calorimeter, two layers of emulsion sheets on top of each other were used: one fixed in space, the other moved by a certain amount in a certain time interval with respect to the other one. The two emulsion film images were then matched and thus the time of the shower passage determined.

\begin{figure}[th]
\centerline{\epsfxsize=3.2in\epsfbox{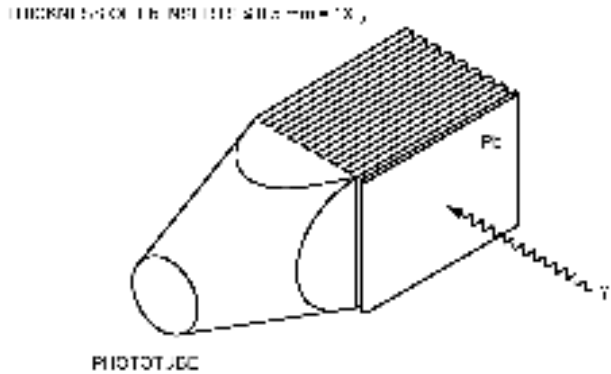}}   
\caption{The schematic drawing of a lucite sandwich counter and a scintillation sandwich counter is shown. \label{fig:inter3}}
\end{figure}

In the 1960s high energy accelerators in the USA and Europe became the main facilities to study high energy phenomena. While charged particles were measured with high accuracy in magnetic spectrometers, the energy of neutral particles like photons, $\pi^{o}$ and neutrons could only be measured by calorimetric means. First successful attempts to measure photon and electron energies with compact electromagnetic sampling calorimeters were made at CALTECH by C. Heusch and C. Prescott\cite{bib:heu}. They studied the electromagnetic shower development of electrons and photons in the energy range of 100 MeV to 5 GeV. For that purpose they built two sandwich counters, one made of plastic scintillator and one of lucite material with lead inserts as absorbers (Fig.\ref{fig:inter3}). With this they were able to exploit and compare the performance of calorimeters based on the ionization loss of shower particles in scintillators and those based on Cerenkov radiation in plastic materials. They also studied sampling fluctutations and shower containment by changing the thickness of the lead inserts. A similar study for a hadron calorimeter was done by the Karlsruhe group under the leadership of H. Schopper motivated by the idea of measuring n-p elastic scattering at CERN\cite{bib:en}.
With the vision that calorimeters will play a role not only in accelerator experiments but also in experiments on board of space missions, R. Hofstadter and his collaborators\cite{bib:hug} developed large homogeneous NaI (Tl) and CsI total absorption calorimeters. Because of their robustness, CsI crystals were particularly suited for space-born gamma ray astronomy. The persuasively good energy resolutions obtained with these calorimeters also made them powerful tools for studying inclusive $\pi^{o}$ production in hadron collisions as well as electron and gamma final states in $e^{+}e^{-}$ collisions at SPEAR. Hofstadter's original ideas were later further developed and realised in the very successful CRYSTAL BALL calorimeter at SLAC, with which many of the charmonium states have been discovered (see chapter 8).

\section{Segmented Calorimeters}

Large charged particle spectrometers were designed for exploring the physics in an energy domain which was made accessible by the new large proton accelerators at Fermilab and at CERN (SPS) in the early 1970s. At that time the usefulness of calorimetric energy measurements to complement the measurements in magnetic spectrometers was not yet widely appreciated. However, the parton picture of hadrons, strongly supported by the SLAC single-arm spectrometer experiment, demanded a search for parton jets, which would for example manifest their presence by a large transverse energy flow $E_{T}$ in deep inelastic hadron-hadron collisions. For this search, segmented calorimeters covering a large solid angle were developed. Although jets could not be unambiguously identified at Fermilab, ISR and SPS energies, they were clearly detected at higher energies in the UA2 and UA1 experiments at the p-\={p} collider at CERN. The extraordinarily successful use of segmented calorimeters in the UA1 \cite{bib:cord} and UA2 \cite{bib:bee} experiments in discovering not only jets but also the intermediate vector bosons in their decay-channels $Z\rightarrow 2\;leptons$ and $W^{\pm}\rightarrow lepton + missing\:E_{T}$ $(E_{T}^{miss})$ made them indispensable tools for future collider experiments searching for the top quark $(t \rightarrow jets + leptons)$, the Higgs $(H \rightarrow \gamma\gamma)$ and SUSY particles $(E_{T}^{miss})$.
\begin{figure}[th]
\centerline{\epsfxsize=3.2in\epsfbox{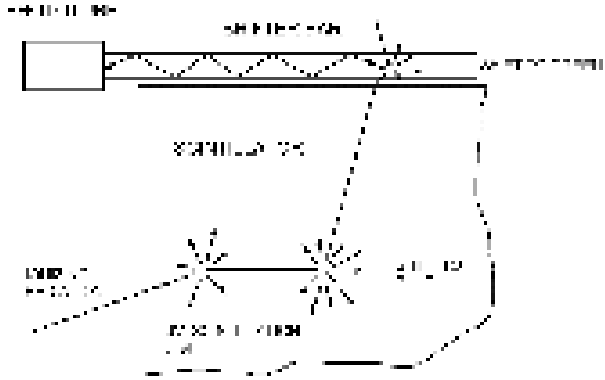}}   
\caption{The principle of a wavelength shifting read-out is shown. \label{fig:inter4}}
\end{figure}
In 1973 neutral currents were discovered by GARGAMELLE at CERN\cite{bib:has}. This important step was decisive for future neutrino experiments. To measure neutral current reactions in neutrino detectors required calorimeters of multiton mass with very large dimensions. One of the first upgrades in this direction was initiated by B. Barish and collaborators\cite{bib:bar} of CALTECH. They improved the calorimetric shower detection capability of their neutrino experiment at Fermilab by introducing a wavelength shifting (WLS) read-out\cite{bib:shu} of their very large-sized scintillator tiles. The principle of the WLS technique is shown in Fig.\ref{fig:inter4}. The primary light in the scintillator is shifted to longer wave length to match the absorption spectrum of the WLS read out bar which is separated by a small air gap from the edge of the scintillator tile. The loss of photoelectrons due to this WLS technique, typically a factor of 10, could be compensated somewhat by using thicker scintillators. However, the main advantage of the WLS technique over the conventional light-guides is its simplicity and the smaller number of read-out channels necessary, which is reflected in the lower costs. By reading the light from at least two opposite sides of a scintillator tile, it was even possible to locate the center of gravity of the shower to within a few cm accuracy.

\section{Segmentation Using Wavelength Shifter Read-out}

The WLS principle, which was first introduced by W. Shurcliff\cite{bib:shu}, further discussed by R.L. Garwin\cite{bib:gar} and later developed by G. Keil\cite{bib:gk} provided the capability to construct tower-structured scintillation counter sampling calorimeters. Tower-structured calorimeters were an essential ingredient for the jet search, since they allowed to trigger on events with a large transverse energy $E_{T}$ and to determine the jet energy as well as the jet size. Two types of WLS systems were developed: WLS sheets and WLS rods (later replaced by WLS fibers).
\begin{figure}[th]
\centerline{\epsfxsize=3.2in\epsfbox{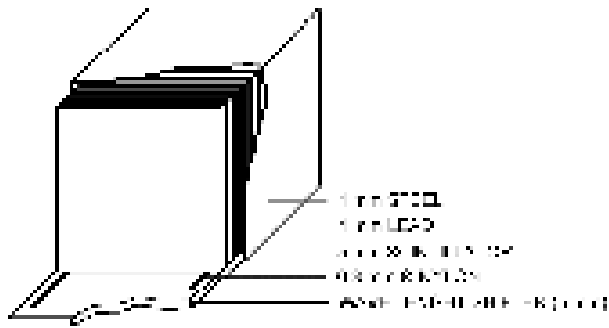}}   
\caption{A  lead scintillator  tower using a wavelength shifter sheet read-out  is shown. \label{fig:inter5}}
\end{figure}

W. Selove\cite{bib:sel} used WLS sheets for his jet search experiment at Fermilab and D. Wegener\cite{bib:hof} developed a fine sampling tower-structured electromagnetic calorimeter for the ARGUS experiment at the DORIS $e^{+}e^{-}$ collider. A lead scintillator tower  using a WLS sheet read-out is shown in Fig.\ref{fig:inter5} from Ref.\cite{bib:hof} . 

\begin{figure}[th]
\centerline{\epsfxsize=3.2in\epsfbox{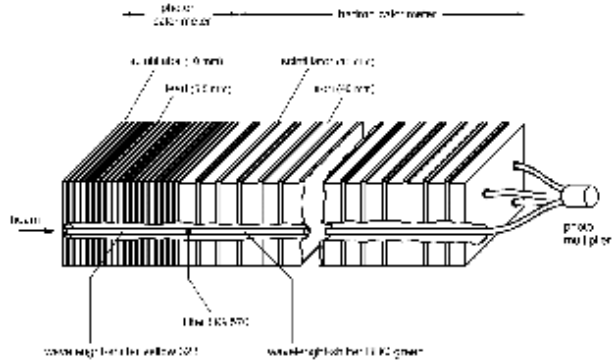}}   
\caption{The wavelength shifter read-out system developed for the NA5 calorimeter is shown. The signals from the electromagnetic and the hadronic part of the calorimeter could be measured separatly. They  were distinguishable  by the yellow (electromagnetic)  and green (hadronic) colour of the wavelength shifter rods. \label{fig:inter6}}
\end{figure}

A novel read-out system using WLS rods\cite{bib:eck} was introduced by the author of this article for the jet search experiment NA5 at the SPS at CERN. By using a different WLS color (yellow) in the electromagnetic part of the NA5 calorimeter\cite{bib:dem} than in the hadronic part (BBQ green), it was possible to guide the light of both colors in one rod to the two photomultipliers which, by using appropriate filters, were sensitive to either yellow or green light. 
This way the energy deposited in the electromagnetic and in the hadronic part of the calorimeter could be measured separately (Fig.\ref{fig:inter6}). The NA5 calorimeter was the first tower-structured scintillator sampling calorimeter in operation at CERN. It is still in use in the NA49 experiment in the North Area of the SPS. On the basis of the first positive experiences with WLS read-outs, UA1 and UA2 at CERN and CDF at Fermilab developed their calorimeters using this principle.

The next step was to introduce WLS-doped optical fibers and scintillating optical fibers for future scintillator sampling calorimeters. However, no company in the fiber-producing industry was interested in developing these special fibers with only slim expectations for a large market. In 1982 D. Treiller, P. Sonderegger and the author of this paper persuaded J. Thevenin at SACLAY to develop scintillating fibers and WLS fibers for calorimetry. Thanks to J. Thevenin, very soon the first tower-structured electromagnetic calorimeter\cite{bib:fes} using WLS fibers doped with K27 (BBQ was not successful, since it cracked during extrusion) could be successfully tested. It was later given the name SHASHLIK. Due to its good energy resolution and its relatively low construction costs, SHASHLIK is being used in DELPHI, HERA-B, PHENIX (RHIC) and LHCB experiments. WLS fiber read-out in hadronic tile calorimeters was further developed by the CDF collaboration at Fermilab. This technique is also applied in massive neutrino calorimeters, like for example in MINOS and in the target tracker in OPERA. 

\begin{figure}[th]
\centerline{\epsfxsize=3.2in\epsfbox{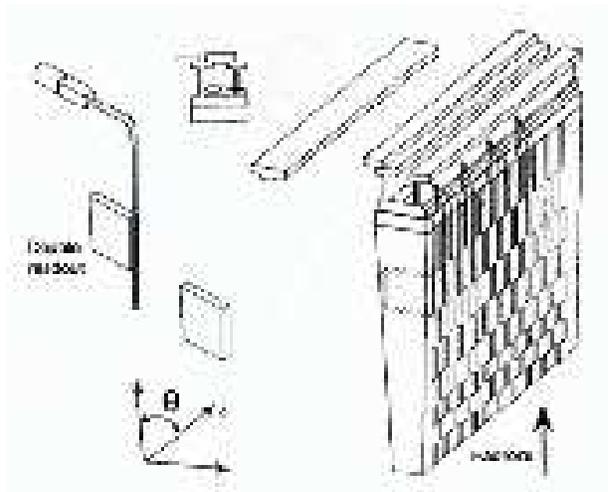}}   
\caption{One of the thirty six ATLAS hadronic barrel calorimeter modules is shown.The longitudinally oriented scintillator tiles are read out by WLS fibers. \label{fig:inter7}}
\end{figure}

O. Gildemeister proposed a novel structure for the ATLAS hadron calorimeter, the so-called TILECAL\cite{bib:gil}. In this structure, the scintillating tiles and the read-out fibers run parallel to the particles (Fig.\ref{fig:inter7}). This design turned out to be cheap and allowed a hermetic construction, projective geometry as well as depth sampling (3 times).

\begin{figure}[th]
\centerline{\epsfxsize=4in\epsfbox{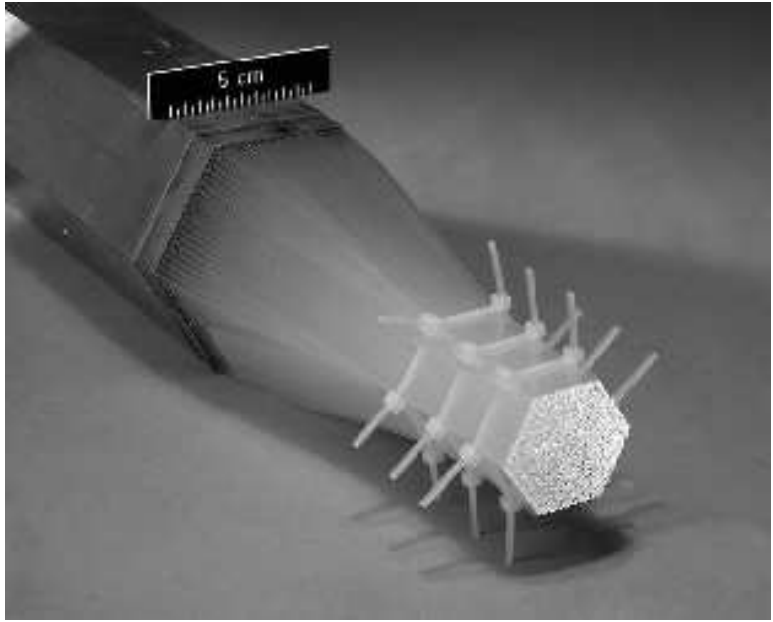}}   
\caption{One of the towers of the SPACAL scintillating fiber calorimeter is shown. \label{fig:inter8}}
\end{figure}

Equally promising were the first tests of J. Thevenin's scintillating fibers used in a prototype calorimeter. This technique was further developed, leading to the SPACAL\cite{bib:aco} calorimeter - at the time a most promising candidate for use in LHC experiments. SPACAL was a compensating (for explanation see chapter 6), homogeneous, high resolution calorimeter with properties very close to homogeneous calorimeters (Fig.\ref{fig:inter8}). However, depth segmentation turned out to be difficult and its production was more expensive than other calorimeters with similar performances. Therefore it finally did not get used in LHC experiments, but its great performance was fully appreciated in smaller-sized experiments like CHORUS, H1 backcalo, AMS-2 and KLOE.

Mainly for heavy ion physics at the CERN SPS there was a demand for a fast and radiation-hard calorimeter to measure forward energy flow, which was used for the determination of the centrality of the collision. P. Gorodetzky\cite{bib:gor} pioneered the development of a quartz fiber/lead calorimeter with a very fast response (6 ns pulse to pulse separation) due to Cerenkov radiation. The radiation resistance of quartz fibers is considerably higher than that of scintillating fibers. However, because of its moderate energy resolution this type of calorimeter is limited in its applications. Quartz fiber calorimeters are employed in the N50, NA52, and CMS (CASTOR) experiments.

\section{Liquid Ionization Chambers}

One of the limiting factors of a sampling calorimeter are the sampling fluctuations and the uniformity of response. To reduce both these factors and still keep the sampling calorimeter compact with minimal dead space between towers, W. Willis and V. Radeka\cite{bib:rad}, in the early 1970s, introduced liquid argon (LAr) as active medium. This technique has proven to be highly successful and has been used in many fixed target and collider experiments (R807/ISR, MARK2, CELLO, NA31, SLD, HELIOS, D0, HERA, ATLAS). The development of the LAr ionization chamber technique was also pioneered by the group of J. Engler\cite{bib:eng} at Karlsruhe. Although fine sampling LAr calorimeters surpass the performance of their scintillator counterparts in many respects, they have the disadvantage that they require cryogenics (boiling temperature of LAr is T = 87 K) and high purity of the liquid.

In order to overcome the cooling problem, J. Engler and H. Keim\cite{bib:km} developed liquid ionization chambers which operate at room temperature. They used tetramethylsilane $Si(CH_{2})_{4}$ (TMS), which has a relatively high electron mobility. TMS and tetramethylpentane (TMP) have the double advantage of operating at room temperature and being hydrogeneous, providing the necessary presupposition for a compensating calorimeter. A disadvantage is the high degree of purification needed. Warm liquid calorimeters have been proposed for the UA1 upgrade (TMP) and are being used in the CASCADE experiment.

\begin{figure}[th]
\centerline{\epsfxsize=4in\epsfbox{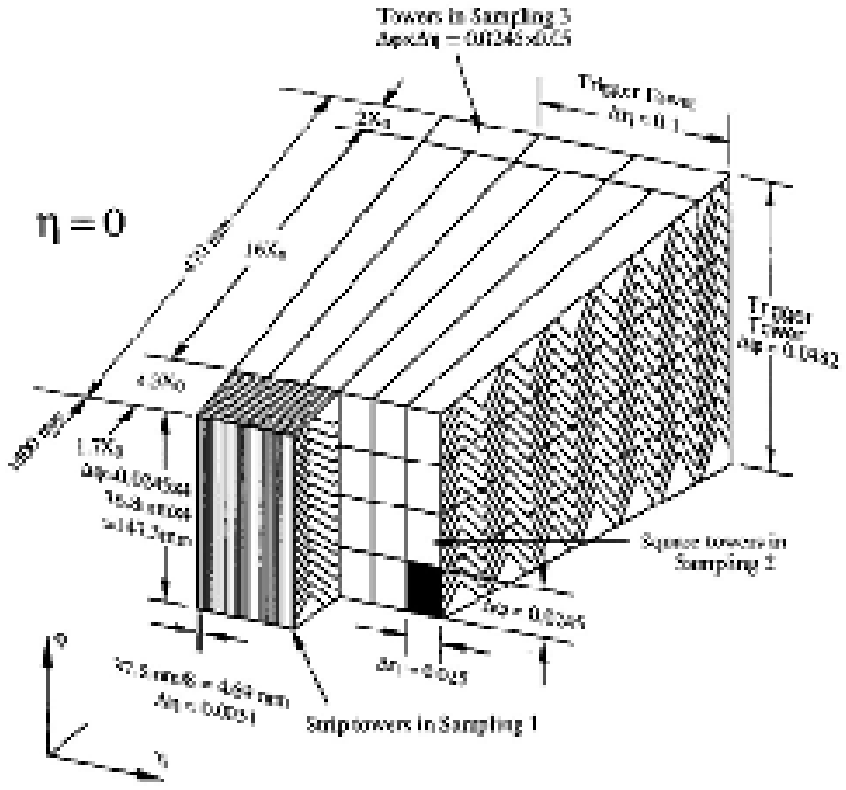}}   
\caption{ATLAS liquid argon electromagnetic calorimeter based on the "accordeon" arrangement. \label{fig:inter9}}
\end{figure}

In 1990 D. Fournier\cite{bib:aub} introduced a novel design for a LAr calorimeter, the so-called "accordeon", which has practically no dead space between towers and provides better uniformity of response, less cabling (signals can be extracted from the front and back face of the calorimeter) and fast signal extraction due to low capacitance. The "accordeon" was adopted as electromagnetic calorimeter for the ATLAS detector (Fig.\ref{fig:inter9}).  In the evaluation of all possible types of calorimeters, CMS chose PbW$O_{4}$ crystals for their electromagnetic calorimeter with the argument of obtaining the best energy resolution, while for ATLAS the better uniformity of response and the better angular resolution with the "accordeon" were decisive.

In the mid-1980s C. Rubbia started the development of a LAr total absorption calorimeter for solar and atmospheric neutrino experiments at the Gran Sasso Laboratory. A 600 ton calorimeter has been realized and is going to be installed in the Gran Sasso Laboratory. The ICARUS collaboration plans to complete the 
calorimeter to 3 kton to perform a long baseline neutrino oscillation experiment using the CNGS beam from CERN. A LKr total absorption calorimeter has been developed by the NA48 collaboration and is in operation to study CP violation in 
$K^{o}_{S,L}\rightarrow\pi^{o}\pi^{o}$ ÷and÷ $K^{o}_{S,L}\rightarrow\pi^{+}\pi^{-}$ channels. At the Paul Scherrer Institute in Villigen a LXe calorimeter is in construction to search for forbidden $\mu\rightarrow\mathrm{e}+\gamma$ decays which, if found, would indicate physics beyond the Standard Model.

\section{Compensating Calorimeters}

Even with minimal sampling fluctuations the energy resolution of hadron calorimeters is mainly limited by large shower fluctuations, which are due to fluctuations in the electromagnetic component ($\pi^{o}$, etc.) of the shower and the invisible energy due to nuclear excitations, muons and neutrinos. It was 
suggested by C. Fabjan and W. Willis\cite{bib:fab} that some of this invisible energy can be recuperated using depleted $^{238}U$ plates as absorber. The energy loss will be compensated by the emission of soft neutrons and gammas in fission processes of the uranium. Measurements supported these ideas and the first compensating $^{238}U$/scintillator calorimeter was employed in the Axial Field Spectrometer at the ISR\cite{bib:gord}.

Compensation turned out to be essential for a correct jet energy determination and therefore was a main consideration for the design of calorimeters for high energy collider experiments. However, there are also other ways to achieve compensation, namely by proper weighing of the electromagnetic and hadronic 
components, as shown by H. Abramowicz et al\cite{bib:abr} for the WA1 calorimeter and W. Braunschweig et al\cite{bib:bra} for the H1 calorimeter.

In 1987, mainly due to the detailed work of R. Wigmans\cite{bib:wig} supplemented by the work of Brueckmann et al\cite{bib:bru}, the compensation effects were fully understood. Wigmans could show that compensation can be achieved with absorber materials other than uranium by tuning the influencing parameters like choosing 
the appropriate sampling fractions. For example, compensation can be achieved by choosing the thickness of the absorber and of the scintillator as follows: U/sci = 1/1, Pb/sci = 4/1, Fe/sci = 15/1. The crucial role of hydrogen in the active material was clearly demonstrated by measurements of the L3 collaboration.

LAr calorimeters do not compensate, due to the missing hydrogen. However, compensation can be achieved by prolonging the charge collection time so that the gammas from neutron capture reactions can be detected or by using the above mentioned weighing procedure.

In order to achieve the best jet energy measurements large compensating calorimeters were built at HERA (ZEUS: $^{238}U$/sci, H1: Pb,steel/LAr) and at the Fermilab collider (D0: $^{238}U$/LAr).

However, compensation degrades the energy resolution for electromagnetic showers considerably. Since the search for the Higgs, with its prominent decay channel $H\rightarrow\gamma\gamma$, is one of the prime goals at the LHC, the ATLAS and CMS experiments have chosen non-compensating calorimeters with the highest energy resolution capabilities for electromagnetic showers.

\section{Other Sampling Calorimeters}

The high segmentation capability and the relatively low cost were the main arguments for the development of gas sampling calorimeters. All four LEP experiments were equipped with such devices. The modest obtainable energy resolution ($\sim 20\% / \sqrt{E}$) for electromagnetic showers with such calorimeters is mainly due to Landau and pathlength fluctuations. In addition, gas sampling calorimeters have to cope with stability problems due to temperature and pressure changes. The disadvantages outweighed the advantages such that they were not further developed.

In 1983 P.G. Rancoita and A. Seidman\cite{bib:ran} introduced silicon detectors for electromagnetic calorimetry. These calorimeters were further developed by the SICAPO collaboration\cite{bib:brb}. Large-sized silicon detectors employing relatively low-resistivity (less expensive) material are appropriate for calorimeters used in experiments requiring compact geometry, fast signal response and operation in strong magnetic fields. Disadvantages are the high cost and the poor radiation resistance of silicon. Nevertheless, they are used in the ZEUS experiment as electron-hadron separator and in the PAMELA experiment as imaging calorimeter and they were employed as luminosity monitor in the OPAL experiment at LEP. Si/W calorimeters are being seriously considered for $e^{+}e^{-}$ linear collider experiments of the next generation.

\section{Crystal Calorimeters}

\begin{figure}[th]
\centerline{\epsfxsize=4in\epsfbox{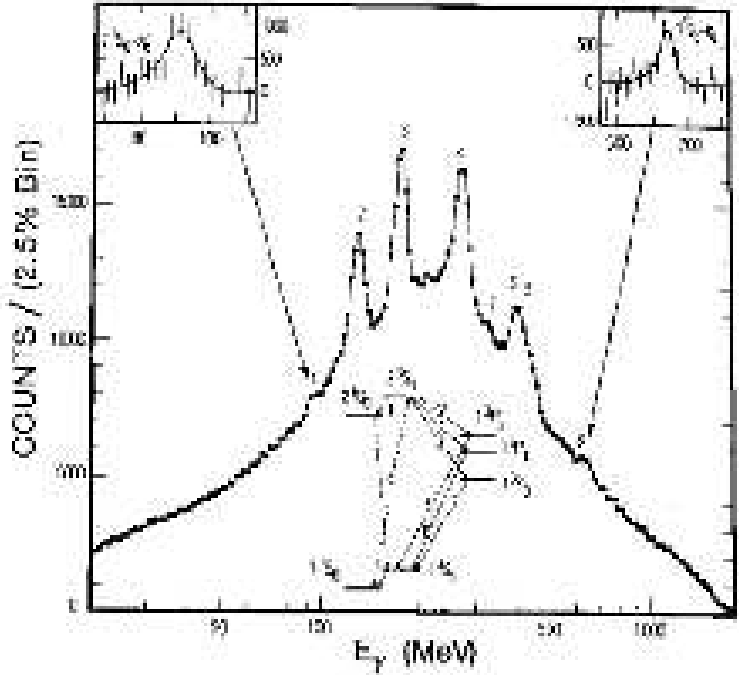}}   
\caption{Charmonium states as measured with the high resolution CRYSTAL BALL calorimeter at SLAC. \label{fig:inter10}}
\end{figure}

Pioneering work in using inorganic Crystals for the detection of high energy particles goes back to V.I.Broser\cite{bib:bros}, H.Kallmann\cite{bib:kall}, R.Moon\cite{bib:moo} and others.
Following earlier developments using NaI (Tl) crystals for physics at $e^{+}e^{-}$ colliders, the CRYSTAL BALL collaboration\cite{bib:ore} at SLAC demonstrated very impressively the discovery potential of a tower-structured high-resolution crystal calorimeter. It made major contributions to the discovery and the better understanding of the charmonium states (Fig.\ref{fig:inter10}). Crystal calorimeters are mainly used for electromagnetic shower detection. In later developments, where the CRYSTAL CLEAR collaboration\cite{bib:crys}  played a leading role, NaI (Tl) crystals have been replaced by denser and non-hygroscopic materials. Among the most popular are: CsI, BGO and PbW$O_{4}$. The production as well as the physical properties of many of these crystals were already studied during World War II, since fluorescent and scintillating crystals were developed to mark firing tanks or guns for airplane attacks at night. The typical energy resolutions of (1.5\% - 3.5\%) / $\sqrt{E}$ obtained with crystal calorimeters have not been reached by sampling devices. Because of their energy resolution and compactness crystal calorimeters have  frequently been applied in collider experiments: L3, CUSB, CLEO II, KTEV, GLAST, BABAR, BELLE and CMS. One of the notable by-products is the use of CsI (Tl) crystals in medical tomography.

\section{Cryogenic Calorimeters}

In 1935 F. Simon\cite{bib:sim} suggested measuring the energy deposited by radioactivity with low temperature calorimeters. He claimed that, with a calorimeter consisting of $1cm^{3}$ tungsten in a liquid helium bath at 1.3 K, one could measure a heat transfer of the order of nWatt, which is about 1000 times more sensitive than the calorimeter of W. Orthmann. The argument is that at low temperatures the heat capacity C of a calorimeter is low and a small energy loss $\Delta\mathrm{E}$ of a particle in the calorimeter can lead to an appreciable temperature increase $\Delta\mathrm{T}=\Delta\mathrm{E}/C$. More recently, the development of cryogenic calorimeters was motivated by the quest for the dark matter in the universe, the missing neutrinos from the sun, the neutrinoless double beta decay and the mass of the neutrinos. First ideas and experimental attempts were discussed at a first workshop on low temperature detectors (LTD1)\cite{bib:pre}, which was held in 1987 at Ringberg Castle in southern Bavaria. More workshops followed every other year, either in Europe or in the USA. A review on cryogenic calorimeters in astro and particle physics can be found in Ref.\cite{bib:pret}.
\begin{figure}[th]
\centerline{\epsfxsize=3in\epsfbox{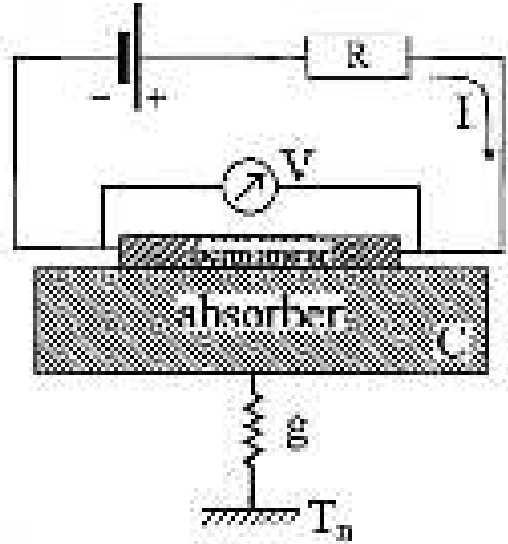}}   
\caption{The principle of a cryogenic calorimeter (bolometer) is shown. \label{fig:inter11}}
\end{figure}

A typical cryogenic calorimeter, a so-called  bolometer, is shown in Fig.\ref{fig:inter11}. It consists of an absorber with heat capacity C, a thermometer and a thermal link with heat conductance g to a heat reservoir with a constant bath temperature $T_{B}$. Particles interacting in the absorber cause a change in the resistance of the thermometer which is measured by a voltage drop V when passing a current I through the thermometer. Cryogenic calorimeters can be made from many different materials including superconductors, a feature which turns out to be very useful for different applications. They can be used as target and detectors at the same time.

\begin{figure}[th]
\centerline{\epsfxsize=4in\epsfbox{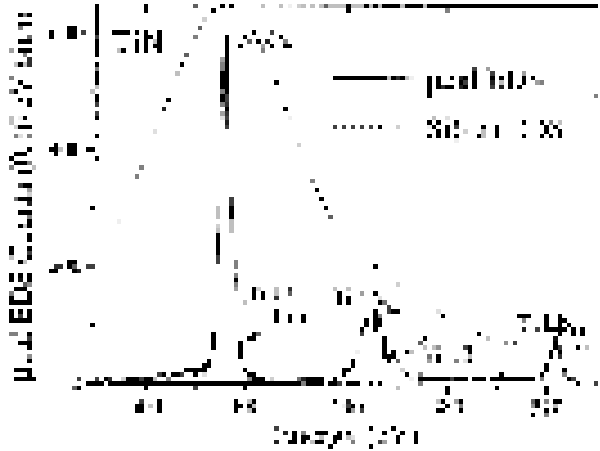}}   
\caption{TiN X-ray spectra obtained with a cryogenic micro-calorimeter (solid line) and with a state of the art Si(Li) solid state device (dashed line) are compared. EDS stands for energy-dispersive spectrometer. TiN is an interconnect and diffusion barrier material used in semiconductor industry. \label{fig:inter12}}
\end{figure}

Most calorimeters used in high energy physics measure the energy loss of a particle in form of scintillation light or ionization. In contrast, cryogenic calorimeters are able to measure the total deposited energy in form of ionization and heat with very high efficiency. This feature makes them very effective in detecting very small energy deposits (order of eV) with very high accuracy. The use of superconductors as cryogenic particle detectors was motivated by the small binding energy $\sim 1\mathrm{meV}$ of the Cooper pairs. Thus, compared to conventional detectors, several orders of magnitude more free charges are produced, leading to a much higher intrinsic energy resolution. In Fig.\ref{fig:inter12} X-ray spectra obtained with a cryogenic micro-calorimeter (solid line) and a state of the art Si(Li) solid state device ( dashed line) are compared. The micro-calorimeter consisting of a Bi absorber and a Al-Ag bilayer superconducting transition edge thermometer has been developed at the National Institute of Standards and Technology (NIST) in Boulder USA \cite{bib:wol}.

The main advantage of cryogenic devices for the direct detection of dark matter particles, so-called WIMPs (weakly interacting massive particles), is their effectiveness in measuring very low energy recoils and the possibility of using a large variety of detector materials. WIMPs can be detected by measuring the nuclear recoil energies in coherent elastic WIMP-nucleus scattering. Depending on the mass of the WIMP and the mass of the detector nucleus, the average recoil varies between eV and keV. The most commonly used WIMP detectors are bolometers shown in Fig.\ref{fig:inter11}. E. Fiorini and T. Niinikoski\cite{bib:fio} pioneered the development of bolometers for measuring neutrinoless double beta decay.

But also superheated superconducting granules (SSG) for dark matter detection\cite{bib:bor} have been developed. This technique was first introduced by H. Bernas et al\cite{bib:bern} for beta radiation detection and later proposed for dark matter and solar neutrino detection by L. Stodolsky and A. Drukier\cite{bib:dk}. The first generation WIMP experiments with cryogenic bolometers with absorber masses of about 1 kg (CDMS, EDELWEISS, CRESST and ROSEBUD) and with SSG of 0.5 kg (ORPHEUS) are in operation. Recent reviews of dark matter detection can be found in Ref.\cite{bib:pretz} and Ref.\cite{bib:gait}. Larger detector masses of more than 10 kg are planned for the future. One of the big advantages of cryogenic calorimeters over conventional WIMP detectors is their capability of active background recognition, which allows to discriminate between background recoils due to electron scattering and genuine nuclear recoils by a simultaneous but separate measurement of phonons and ionization (or photons) in each event. This is possible since, for a given deposited energy, the ionization generated by nuclear recoils is smaller than that generated by electrons. This feature, which was first pointed out and further developed by B. Sadoulet\cite{bib:sad} increases the sensitivity for WIMP detection considerably. A similar idea using scintillating crystals as absorbers was introduced by L. Gonzales-Mestres and D. Perret-Gallix\cite{bib:gon} and further developed by ROSEBUD and CRESSTII.

Mainly for applications other than WIMP detection, calorimeters on the basis of superconducting tunnel junctions (STJ) have also been studied. Cryogenic cameras\cite{bib:pea} consisting of STJ pixel arrays provide the astronomers with a powerful tool to observe very faint and distant objects and to determine their distance via red shift. Other research areas have also benefited from these cryogenic detector developments, such as x-ray spectroscopy in astrophysics, mass spectrometry of large molecules (DNA sequencing) as well as x-ray microanalysis for industrial applications.

\section{Detection of Extraterrestrial Neutrinos}

In 1968 R. Davis\cite{bib:dav} pioneered the detection of extraterrestrial neutrinos and started a new field of neutrino astronomy. In his chlorine detector buried in the Homestake mine in South Dakota he was the first to detect neutrinos from the sun. The neutrino flux, however, turned out to be lower than expected from the standard solar model. His findings were later confirmed by many experiments, such as BAKSAN, GALLEX, GNO, KAMIOKANDE, SUPERKAMIOKANDE. However, the missing neutrinos from the sun still remained a puzzle until 2002, when the SNO experiment\cite{bib:anm} confirmed a long-presumed hypothesis, namely that neutrinos have a mass and as a result of this they oscillate, i.e. they change flavour on their way from the sun to the earth. Anti-electron neutrino oscillations have also been measured with reactor neutrinos in the  Kam-Land experiment\cite{bib:eug}  supporting the SNO findings.

The discovery of atmospheric muon neutrino oscillations in 1998 by the SUPERKAMIOKANDE water Cerenkov detector\cite{bib:fuk} was made possible due to the intuition and initiative of M.Koshiba and Y.Totsuka, who were responsible for the construction of KAMIOKANDE and SUPERKAMIOKANDE. These oscillations have later also been observed with the MACRO detector in the Gran Sasso Laboratory (Italy) and have recently been confirmed in the K2K experiment\cite{bib:nak} by sending accelerator born neutrinos to SUPERKAMIOKANDE. 

In order to look for extraterrestrial neutrinos, K. Lande\cite{bib:lan} and collaborators installed a water Cerenkov detector in the Homestake mine adjacent to the R. Davies experiment. The detector consisted of 7 water tanks, each viewed by 4 PMs on opposite sides of the tank. On January 4, 1974 the detector signalized an event which the experimenters interpreted as a possible antineutrino burst. Unfortunately, such events do not happen very often and a confirmation is practically impossible unless they are also witnessed by other experiments. If for nothing else, these findings wakened the curiosity of researchers to look for more of these events with larger and better detectors. Thus it was around this time that F.Reines from Irvine, S.Miyake from INS Tokyo, J.G.Learned and V. Peterson from Hawaii and others started to think about a deep sea water neutrino telescope near the coast of Hawaii which later became a project called DUMAND\cite{bib:rob}. The idea to use the ocean for Cerenkov light detection goes back to M.A.Markov\cite{bib:mar}. Unfortunately DUMAND was  discontinued, but it paved the way for the next generation of cosmic neutrino Cerenkov detectors: NESTOR, ANTARES and NEMO in the Mediterranean sea as well as AMANDA and ICECUBE in the Antarctic ice.

Massive calorimeters like NUSEX in the Mont Blanc tunnel, BAKSAN in Russia, IMB in the Morton-Thiokol salt mine and KAMIOKANDE in Japan, originally designed for proton decay experiments, were, on 17th February 1987, witness to an extraterrestrial neutrino burst, which originated from a nearby supernova explosion (SN1987A). This was the first time neutrinos from a supernova explosion were detected with terrestrial calorimeters. This happy event encouraged all those who had already made plans to set new trends in neutrino astronomy.

\section{The Atmosphere as Calorimeter}

At sea level our atmosphere measures 1032 $g/cm^{2}$. Very high energy particles from outer space trying to traverse our atmosphere would find material in front of them which corresponds to 28 radiation lengths and 16.6 collision lengths. Particles from cosmic origin like hadrons, photons or neutrinos interact with air nuclei, producing secondaries that in turn collide with air atoms, leading to extensive air showers (EAS). The most numerous particles in EAS are electrons. Electrons traversing the atmosphere produce Cerenkov light, which is directed along their path. On their way, they can also excite metastable energy levels in atmospheric molecules which, after a short relaxation time, emit a characteristic fluorescence light, which peaks at wavelenghts from 330 to 450 nm. In contrast to the Cerenkov light the emitted fluorescent light is isotropic. The emitted Cerenkov and fluorescence light are proportional to the EAS energy. Thus properly instrumented, the atmosphere can be utilized as a huge calorimeter.

\begin{figure}[th]
\centerline{\epsfxsize=4in\epsfbox{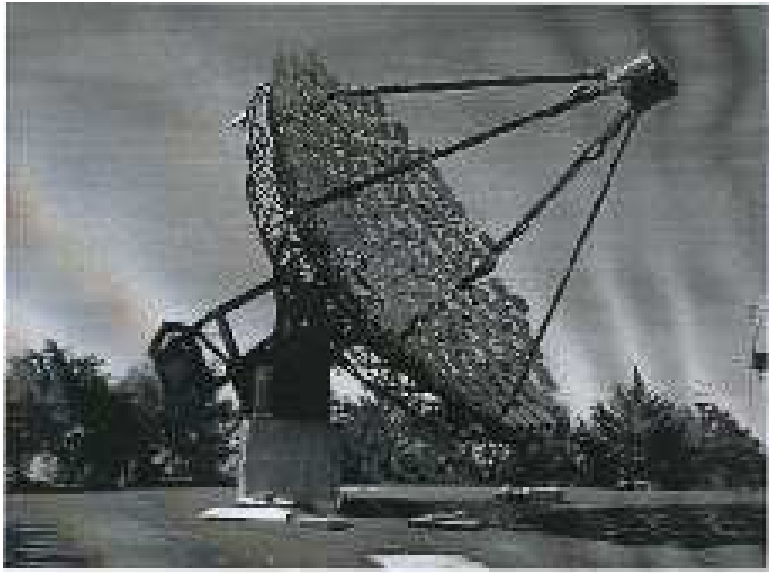}}   
\caption{The Whipple Observatory 10 m imaging atmospheric Cherenkov telescope. \label{fig:inter13}}
\end{figure}

It was P. Blackett\cite{bib:bla} who, in 1948, suggested to detect the Cerenkov light in the atmosphere caused by penetrating cosmic particles. A. Chudakov and collaborators\cite{bib:chu} were the first to apply this idea in 1962 to detect celestial $\gamma$-rays. Atmospheric Cerenkov light detection was also pioneered by the Whipple telescope on Mount Hopkin, Arizona, in 1987(Fig.\ref{fig:inter13}). It consisted of a 10m-diameter mirror dish and a pixel array of photon detectors\cite{bib:cw}.

\begin{figure}[th]
\centerline{\epsfxsize=4in\epsfbox{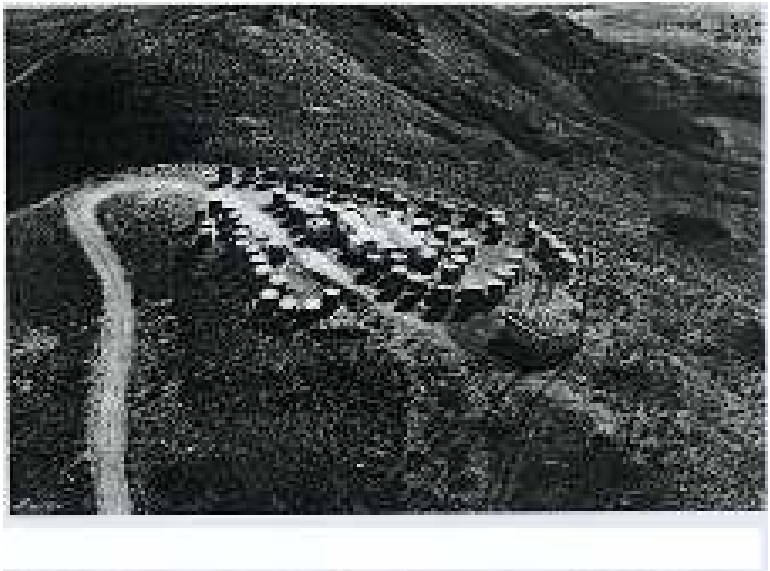}}   
\caption{Aerial view of the Fly's Eye I detector on top of the Little Granite Mountain in Utah. \label{fig:inter14}}
\end{figure}

This technique provided information on the direction, shape and energy of the shower as well as on the type of primary particle (cosmic hadrons produce 2 times less Cerenkov light than gammas of the same energy). The Whipple telescope was one of the most powerful early instruments which made major contributions to the study of high energy gamma rays in the energy range of several hundred GeV to several TeV. However, this technique was limited to a relatively small fiducial area yielding low event rates for high energy showers as well as too poor energy resolutions. In a further step the Fly's Eye detector\cite{bib:rmb} was built to overcome these deficiencies. The Fly's Eye detector consists of an array of spherical mirrors with a cluster of PMTs mounted in the focal plane of each mirror. It records the fluorescence light, which is caused by ultra high energy cosmic ray showers in the atmosphere. The Fly's Eye observatory in Utah consists of two detector stations (Fly's Eye I and II) situated 3.3 km distant from each other (Fig.\ref{fig:inter14}). Fly's Eye I is equipped with 67 detector arrays and Fly's Eye II with 8 detector units with 120 eyes. Events more than 20 km away from the observatory could be detected, giving rise to a very large fiducial area of about 100 $km^{2}$ sr. The simultaneous observation from both Fly's Eye stations allows a stereoscopic reconstruction of an event. From the observed shower profiles the total energy can be derived. However, the knowledge of the energy scale is still a central issue for atmospheric calorimeters. The original Fly's Eye detector has now been replaced by a new facility called HiRes, which is sensitive in the energy domain of the Greisen-Zatsepin-Kuzmin (GZK) cut-off above $10^{20}$ eV . Air showers in this energy range were also observed by the AGASA array in Akeno, Japan. For some time HiRes and AGASA yielded conflicting results in the energy range where the GZK cut off is expected to occur. However, the statistical and systematical uncertainties in both experiments are still very large.  Experiments like AUGER and EUSO/OWL (see below) with better statistics may be able to settle this discrepancy in the future.

The new generation of atmospheric Cerenkov telescopes has the potential to provide images of high energy gamma sources in the energy range of around 0.1 to 10 TeV with better resolution. Two telescopes of the High Energy Stereoscopic System (HESS) in central Namibia and MAGIC in La Palma (Canary Islands) are already taking data. The HESS facility ( the acronym is also an homage to Victor Hess, the discoverer of cosmic radiation) consists of 4 telescopes each equipped with a 107$m^{2}$ mirror area and a camera of 960 PMT's.  MAGIC has two telescopes each with a 239$m^{2}$ mirror and a 577 PMT camera. By 2006 the US project VERITAS  plans to have 4 Cerenkov telescopes in operation.

 Presently under construction is the P. Auger observatory (AUGER) in Malargue in Argentina. It will consist of 1600 water Cerenkov detectors and 4 fluorescence "eyes" spread over an area of 3000 $km^{2}$. AUGER will measure the energy and arrival direction of UHE cosmic rays with energies in excess of $10^{19}$ eV.

For the future, an exploratory mission probing the extremes of the universe using the highest energy cosmic rays and neutrinos is planned. It is called EUSO/OWL (EUSO for Extreme Universe Space Observatory) and will orbit the earth in 500 km height and observe an area of about $3\cdot10^{5}km^{2}\mathrm{sr}$ of the earth's atmosphere, being able to detect several thousand air showers above $10^{20}$ eV per year. With this sensitivity the experimenters aim to systematically study the energy spectrum around the GZK cut-off. They also hope to be able to detect relic Big Bang neutrinos through the $Z_{0}$-resonance absorption of cosmic neutrinos with energies $>10^{21}$ eV.

\section{Conclusions}

The development of novel detectors is essential for the exploration of new domains in physics. Calorimeters are a very good example for this. Major discoveries, like neutral currents ( GARGAMELLE), quark and gluon jets ( SPEAR, UA2, UA1 and PETRA), W,Z bosons ( UA1 and UA2), top quark ( CDF and D0), neutrinos from the supernova explosion SN1987A ( NUSEX, IMB, KAMIOKANDE and BAKSAN), atmospheric neutrino oscillations ( SUPERKAMIOKANDE) and solar neutrino oscillations ( SNO) were made with detectors employing calorimeters. The future will provide enough challenges for young people with imagination to work in this field.

\section*{Acknowledgments}

I would like to thank A. Ereditato, P.Grieder, S.Janos, S.Kabana and U. Moser for the critical reading of the manuscript and usefull discussions.

\end{document}